# Implementations of Cooperative Games Under Non-Cooperative Solution Concepts[*]

Justin Chan[†]

**Abstract**

Cooperative games can be distinguished as non-cooperative games in which players can freely sign binding agreements to form coalitions. These coalitions inherit a joint strategy set and seek to maximize collective payoffs. When the payoffs to each coalition under some non-cooperative solution concept coincide with their value in the cooperative game, the cooperative game is said to be implementable and the non-cooperative game its implementation. This paper proves that all strictly superadditive partition function form games are implementable under Nash equilibrium and rationalizability; that all weakly superadditive characteristic function form games are implementable under Nash equilibrium; and that all weakly superadditive partition function form games are implementable under trembling hand perfect equilibrium. Discussion then proceeds on the appropriate choice of non-cooperative solution concept for the implementation.

## 1. Introduction

From the outset, cooperative and non-cooperative games appear to offer entirely distinct models of collective behavior, but in fact their difference in specification stems solely from a key change of assumption. Cooperative games are arguably best understood as strategic games in which players may costlessly sign binding agreements to form coalitions that aim to maximize their collective payoffs. Unlike in non-cooperative games, these agreements to cooperate bind under external enforcement rather than self-enforcement. That is, they may be enforced by contract law issued by the state, while those in non-cooperative games must rely on credible threats issued by the individual.

Unsurprisingly, this modification significantly widens the scope of potential cooperation amongst players, which in turn allows for a remarkable simplification in the specification of cooperative games. Instead of requiring strategy sets and payoff functions, (zero-normalized) cooperative games are uniquely defined by a player set and a worth function mapping coalitions to a value that represents what players within that coalition can feasibly achieve through working together. Here and in subsequent references to cooperative games, we will assume transferrable utility. One can therefore interpret the values of each coalition in monetary terms with players able to aggregate and divide what they receive across members within their coalition.

---

[*] I am deeply thankful to Eric Maskin and Jerry Green for providing extensive support and suggestions on this paper. I appreciate the feedback from John Geanakoplos, Yannai Gonczarowski, Faruk Gul, Ryoto Iijima, Shengwu Li, David Pearce, Matthew Rabin, Debraj Ray, Slavik Sheremirov, Tomasz Strzalecki, and Richard Zeckhauser. An early draft of this paper was developed while I was an undergraduate at Harvard College and was submitted as my senior thesis. During that stage, the project received funding from Harvard University.
[†] Department of Economics, New York University. Email: j.chan@nyu.edu.

For generality, the worth functions considered here can assign coalitions different values depending on the coalitional structure of the other players. That is, we define the worth function's domain as the set of all *embedded coalitions* – tuples consisting of the coalition and the partition of the player set to which it belongs. These worth functions are more specifically known as *partition functions*, and the games they induce are in *partition function form* (Thrall and Lucas, 1963). When the values assigned to coalitions are independent of the partition, the worth functions are the more commonly used *characteristic functions* (Von Neumann and Morgenstern, 1944). For notational convenience, we omit the partition when detailing the input to characteristic functions. In this sense, characteristic function form games are a strict subset of partition function form games.

It is important to emphasize that the worth function should be thought as a summary statistic of some underlying non-cooperative game, capturing the payoff outcomes that coalitions can feasibly achieve without needing to specify the strategies needed to generate such outcomes. For well-defined cooperative games, one can then outline plausible axioms regarding collective rationality and/or normative concerns to generate an appropriate cooperative solution concept. We designate this process of first obtaining a cooperative game from a non-cooperative primitive as the *exogenous coalition formation approach*. The term exogenous refers to the fact that we hypothesize that every coalition can form and that we speculate as to what the consequences of this formation may be. This was the approach adopted by von Neumann and Morgenstern (1944) as part of the early inception of cooperative game theory.

In contrast, a subsequent line of literature that we refer to as the *endogenous coalition formation approach*, characterized by papers such as Scarf (1971) and Ray and Vohra (1997), derives the cooperative solution concept directly from the non-cooperative game, thereby removing the need to solve for the worth function explicitly. One obvious benefit of this approach is the concision from focusing solely on what can feasibly be achieved by coalitions that can feasibly form. Similarly, this approach may yield cooperative solutions in certain cases where the exogenous approach fails to do so. In particular, the values of some coalitions may be indeterminate under the exogenous coalition formation approach, in which case no well-defined worth function will exist for the whole game. But as long as these coalitions have no incentive to form in the first place, we may remain able to solve for a solution under the endogenous approach.

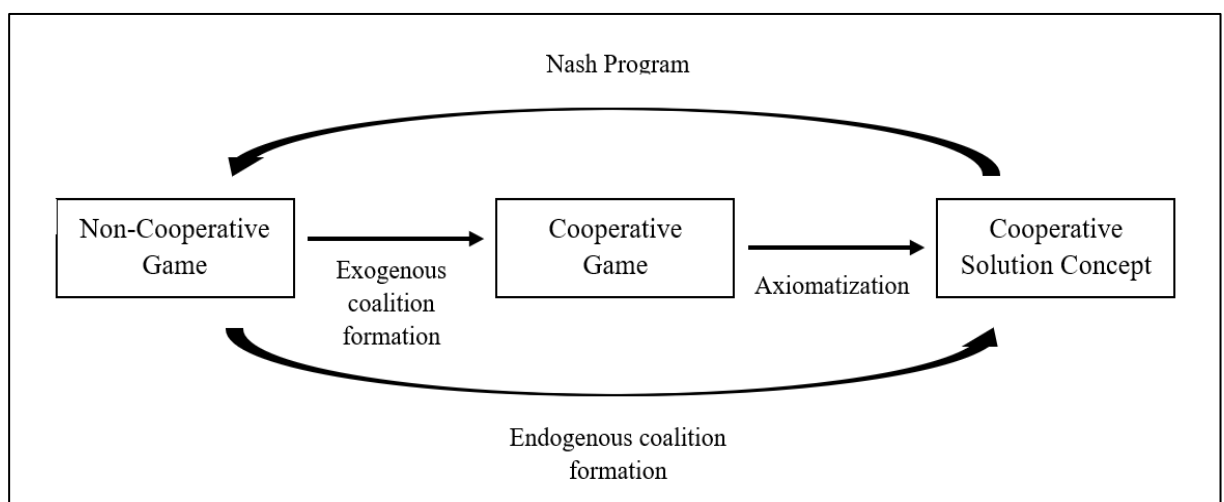

*Figure 1: A Comparison of the Different Approaches to the Simultaneous Treatment of Non-Cooperative and Cooperative Game Theory*

On the other hand, first deriving a worth function from a non-cooperative game provides us with a simpler model to work with, from which we can then deliberate as to what the appropriate cooperative solution should be. Having this flexibility matters as there may be little consensus ex ante on the set of axioms to apply. As long as the cooperative game captures all that is relevant in the non-cooperative game, we can use the worth function to evaluate different cooperative solutions against each other with greater computational efficiency.[1] Furthermore, some cooperative solution concepts may not be directly implementable by a non-cooperative game, and yet they remain useful as benchmarks for stability or fairness. And even if an implementation does exist, it may impose overly strong assumptions on the sequencing and reversibility of coalition formation beyond what the model can justify. Meanwhile, as this paper will demonstrate, large classes of cooperative games can be implemented by a non-cooperative game under quite general and sparing requirements. For these reasons, this paper favors more the spirit of the exogenous approach in its attempt to establish non-cooperative foundations for particular cooperative games.[2]

Nevertheless, there has not historically been a consensus amongst game theorists as to precisely what this micro-founding should entail. As mentioned previously, the simultaneous treatment of non-cooperative and cooperative games began with Von Neumann and Morgenstern (1944), which focused primarily on zero-sum games. It was assumed that any coalition formed from the player set would give rise to its complement. The value of each coalition would be their maximin payoff – or equivalently in this case, their Nash equilibrium payoff – from the resulting two-player game. However, one obvious setback of this framework is that such micro-foundations neglect the many more economic interactions that offer players general-sum payoffs.[3]

Attempts to address this include Aumann and Peleg (1960), which considered general-sum games under maximin outcomes but removed the assumption that strategic interaction be limited to a coalition and its complement. Worth functions derived from such games later became known as α-characteristic functions (Shubik, 1982).[4] But as will be discussed later in the paper, outside of zero-sum games, maximin outcomes are a poor choice of solution concept, being overly pessimistic in predicting what players in a non-cooperative setting can feasibly achieve. In fact, they may result in cooperative solutions that intuitively are unfair given the context, despite still satisfying a given axiomatization of fairness.

---

[1] Classic papers that derive cooperative solution concepts taking the cooperative game as given include Shapley (1953), Schmeidler (1969), and Shenoy (1979). More recent works that take a partition function form game as given include Hafalir (2007) and Grabisch and Funaki (2012).

[2] This paper also shares much the same spirit as the Nash Program insofar as both take the cooperative game/solution concept as given and try to find a corresponding non-cooperative implementation (see Figure 1). However, as this paper is not primarily concerned with cooperative solution concepts, our coverage of the Nash Program will be brief. Noteworthy papers in the field include Chatterjee et al. (1993), Perry and Reny (1994), and Serrano and Vohra (1997). Interested readers should also consult a more recent summary of the history of the Nash Program in Serrano (2021).

[3] In von Neumann and Morgenstern's treatment of *n*-player general-sum games (1944), it was suggested that a fictitious player be introduced such that the resulting *(n+1)*-player game was zero-sum. This fictitious player could not play strategies that influenced the non-cooperative outcome but could compensate players for their cooperation and thus could affect the cooperative outcome of the game. They derived the *extended characteristic function* by applying the same procedure for zero-sum games of equating coalitions' values with their maximin payoffs. Removing all coalitions containing the fictional player from the domain of this function gave the *restricted characteristic function*. Neither worth function gained much traction as an analytic tool in cooperative game theory.

[4] A coalition $S$ was considered α-effective for some payoff $u_S$ if there existed a strategy for $S$ that guaranteed a payoff of $u_S$, regardless of the strategies played by its complement. If all coalitions were α-effective, the α-characteristic function would be well-defined and would set the value of each coalition $S$ as the largest such $u_S$. Aumann and Peleg (1960) also laid the foundations for the β-characteristic function, analogously derived using minimax outcomes instead. Although the authors hypothesized that the β- approach would show more promise as a cooperative concept, the α-characteristic function superseded it in popularity.

It was not until the popularization of the endogenous approach by Scarf (1971) followed by seminal papers such as Hart and Kurz (1983) and Ichiishi (1981), later also Zhao (1992), that the domain of consideration was broadened to general-sum games under Nash equilibrium.[5] What coalitions could feasibly achieve was presented as an equilibrium outcome in which play was cooperative within coalitions and non-cooperative across coalitions. This mathematical formalization was later picked up by papers such as Chander and Tulkens (1997), which assumed that the value of a coalition $S$ would be their Nash equilibrium payoff when playing against the remaining players acting as singleton coalitions. Worth functions derived in this way were named $\gamma$-characteristic functions.

But developments in the exogenous approach still remained limited in that they failed to accommodate partition function form games more broadly. Independence of the worth function and the player partition is a strong assumption to make as many scenarios may exhibit externalities from coalition formation or dissolution. That is, in a wide variety of situations, coalitional payoff can be additionally influenced by the extent of cooperation amongst players outside the coalition. These situations include multilateral trade (Yi, 1996), imperfect competition (Ray and Vohra, 1999), and bargaining over the use of common access resources (Funaki and Yamato, 1999), whereby players free-ride off of a more profitable environment when coalitions are few in number. The specification of partition function form games affords us sufficient flexibility in modeling such externalities.[6]

Any attempt to use maximin outcomes as the basis for a partition function with externalities fails since the value prescribed to a coalition would be the best collective payoff it can achieve regardless of the behavior of the other coalitions. By definition, this excludes the representation of externalities. Meanwhile, any assumption regarding how other players react to the formation of a coalition restricts the scope of consideration to particular partitions of the player set, thus potentially obscuring the prevalence of externalities.

With these concerns in mind, this paper adopts the following convention for the derivation of cooperative games from general-sum games. For convenience, we consider games in normal form. Given such a game, we can partition the player set into coalitions and adjust the strategy sets and payoff function accordingly so as to reflect the fact that players within any given coalition now play strategies as part of some joint action to maximize the collective payoff. We introduce the term *composite game* to designate the resulting games; they are *composite* in the sense that we may decompose the coalitions, strategies, and payoffs prior to aggregation to restore the original normal-form game.

Evidently, the number of composite games that can be constructed from a strategic game equals the number of partitions of its player set. We require that every composite game must have a unique solution or, in the case of multiple solutions, that payoffs to the players are identical across solutions. This ensures that there is no ambiguity as to what coalitions can feasibly achieve and that each coalition's value is well-defined. For consistency, the same non-cooperative solution concept must apply across all composite games. We define the worth function by assigning to each embedded coalition its solution payoff in the corresponding composite game. We say that the resulting cooperative game is *implementable* under the pre-specified non-cooperative solution concept. Or equivalently, the non-cooperative solution concept admits an *implementation* (in this case, the underlying normal-form game) for the cooperative game.[7]

---

[5] See Ray (2007) for a more comprehensive account of the endogenous coalition formation approach and Yi (2003) for a more specialized treatment on coalition formation in partition function form games.
[6] For a thorough treatment of partition function form games and their solutions from both an endogenous and exogenous coalition formation perspective, see Kóczy (2018).

This paper provides novel theorems regarding the existence and construction of implementations for particular classes of cooperative games. Strictly superadditive partition function form games are implementable under Nash equilibrium, as are weakly superadditive characteristic function form games. Moreover, strictly superadditive partition function form games are implementable under certain generalizations and refinements of Nash equilibrium, namely rationalizability and trembling hand perfect Nash equilibrium respectively. In fact, this paper further proves that trembling hand perfect Nash equilibrium admits implementations for weakly superadditive partition function form games, an even larger class of cooperative games. This paper ends with a discussion of the appropriate choice of non-cooperative solution concept, using the example of oligopolistic competition to illustrate how the same implementation under different solution concepts can yield vastly different cooperative outcomes.

## 2. Preliminaries

### 2.1 Notation and Definitions for Non-Cooperative Games

Take a normal-form game $\Gamma = (N, (X_i)_{i \in N}, (u_i)_{i \in N})$ where $N = \{1, 2, \ldots, n\}$ is the player set. Let $\Pi(N)$ denote the set of all partitions of $N$ and $[N] = \{\{1\}, \{2\}, \ldots, \{n\}\}$ denote the finest partition of this player set. We refer to $\{N\}$ as the partition containing the grand coalition. For any partition $\pi \in \Pi(N)$, we denote the size or cardinality of the partition as $|\pi|$. For any two partitions, $\pi, \pi' \in \Pi(N)$, we say that $\pi'$ is a *refinement* of $\pi$, if for all coalitions $S \in \pi$, there exists a subset of $\pi'$ denoted $\mathcal{P}$ such that $\bigcup_{T \in \mathcal{P}} T = S$. We say that $\pi'$ is a *strict refinement* of $\pi$ if $\pi'$ is a refinement of $\pi$ and $\pi \neq \pi'$.

Let $X_i$ denote the finite set of pure strategies of player $i$. For any coalition $S \in \pi \in \Pi(N)$, we let $X_S = \times_{i \in S} X_i$ denote their joint pure strategy set. For any given strategy $\sigma$, we use subscript notation to denote the player to whose strategy set it belongs (this may be a coalition) and superscript notation as an index within that strategy set. We identify any dominant strategies with the asterisk superscript.

We adopt the convention $(\sigma_S, \sigma_{\pi \setminus \{S\}}) \in \times_{S \in \pi} X_S$ to denote the strategy profile consisting of the joint action played by coalition $S$ in the partition $\pi$, holding fixed the strategies played by the other coalitions. We denote $u_S(\sigma_S, \sigma_{\pi \setminus \{S\}}) = \sum_{i \in S} u_i(\sigma_S, \sigma_{\pi \setminus \{S\}})$ to be the payoff to coalition $S$ at this strategy profile.

For convenience, we also define the projection functions $proj_i : \times_{i \in N} X_i \to X_i$ and $proj_S : \times_{S \in \pi} X_S \to X_S$ as a means of extracting the actions played within a given strategy profile by player $i$ and coalition $S$ respectively. Thus, $proj_i(\sigma_1, \ldots, \sigma_n) = \sigma_i$ and $proj_S(\sigma_S, \sigma_{\pi \setminus \{S\}}) = \sigma_S$.

---

[7] Using this terminology, we can summarize the results of cited works as follows: maximin outcomes admit a zero-sum implementation to characteristic function form games (von Neumann and Morgenstern, 1944); α-characteristic function form games are implementable under maximin outcomes (Shubik, 1983); and γ-characteristic function form games are implementable under Nash equilibrium (Chander and Tulkens, 1997).

**Definition 1: Composite Games**

Given $\Gamma$ and some partition of the player set $\pi \in \Pi(N)$, we denote the corresponding *composite game* as $\Gamma_\pi = (\pi, (X_S)_{S \in \pi}, (u_S)_{S \in \pi})$.

Note that $\Gamma$ is a degenerate composite game of itself insofar as $\Gamma = \Gamma_{[N]}$.

**Definition 2: Composite Equilibria**

Under a pre-specified non-cooperative solution concept, we say that the strategy profile $x \in \times_{i \in N} X_i$ is a *composite equilibrium* of $\Gamma$ if there exists a partition $\pi \in \Pi(N)$ such that $x$ solves the composite game $\Gamma_\pi$.

That is, the strategies in a composite equilibrium of $\Gamma$ played jointly or otherwise form the solution to a composite game of $\Gamma$. In fact, the same strategy profile may solve multiple composite games of $\Gamma$.

The following definition provides insight into sufficient means of identifying composite equilibria when the solution concept is Nash equilibrium.

**Definition 3: Potentially Strictly/Weakly Dominant Strategies**

Given some strategy profile $x \in \times_{i \in N} X_i$, we say that player $i$'s strategy $\sigma_i = \text{proj}_i(x)$ is *potentially strictly/weakly dominant in $x$* if there exists a coalition $S$ containing player $i$ such that the joint strategy $\times_{j \in S} \sigma_j$, where $\sigma_j = \text{proj}_j(x)$ for all $j \in S$, is the strictly/weakly dominant strategy of coalition $S$. That is, $\times_{j \in S} \sigma_j = \sigma_S^*$.

As the following lemmata demonstrate, we may be able to construct payoff functions $(u_i)_{i \in N}$ and strategy sets $(X_i)_{i \in N}$ of a game $\Gamma$ such that the composite equilibria of $\Gamma$ are strategy profiles in which all strategies are potentially strictly dominant and vice versa. In doing so, we will first need to introduce two functions.

Let $\kappa: \times_{i \in N} X_i \times 2^N \to \mathbb{N}$ be the function that maps a strategy profile and subset of the player set to the number of potentially dominant strategies played by players within this subset in the strategy profile. And let $\rho_x: N \to 2^N$ be the function defined by

$$\rho_x(i) = \begin{cases} S & \text{if } i \in S \text{ and } \exists \sigma_S^* \in X_S \text{ s.t. } \times_{j \in S} proj_j(x) = \sigma_S^* \\ \emptyset & \text{otherwise} \end{cases} \quad (1)$$

For a fixed strategy profile $x$, we interpret $\rho_x$ as mapping a player to its corresponding coalition if the player's strategy in $x$ is potentially dominant and to the empty set otherwise. Note that it need not be the case that $\rho_x$ is well-defined for every strategy profile. This occurs when there exist distinct $S, T \in 2^N$, some player $i \in S \cap T$, and a strategy profile $x$ such that $\times_{j \in S} proj_j(x) = \sigma_S^*$ and $\times_{j \in T} \text{proj}_j(x) = \sigma_T^*$. But when $\rho_x$ is well-defined, we can infer from (1) the following relationship between $\kappa$ and $\rho_x$:

$$\kappa(x, S) = |S| - \sum_{i \in S} \mathbf{1}(\rho_x(i) = \emptyset) \quad (2)$$

From (2), we can easily see that for all partitions of the player set $\pi \in \Pi(N)$:

$$\kappa(x, N) = \sum_{S \in \pi} \kappa(x, S) \quad (3)$$

**Lemma 1:** *Suppose a game $\Gamma = (N, (X_i)_{i \in N}, (u_i)_{i \in N})$ is such that for all coalitions $S \in 2^N$, there exists a strictly dominant strategy $\sigma_S^*$ in their strategy set $X_S$. Then, for any composite equilibrium of $\Gamma$ under Nash equilibrium denoted $x$, $\kappa(x, N) = |N|$.*

**Proof:** If all coalitions have a strictly dominant strategy, then all composite games are dominant-solvable and admit a unique Nash equilibrium in which these strategies are played. It follows that all composite equilibria consist only of potentially dominant strategies. ∎

**Lemma 2:** *Suppose a game* $\Gamma = (N, (X_i)_{i \in N}, (u_i)_{i \in N})$ *is such that* $\rho_x$ *is well-defined for all strategy profiles. Then, a strategy profile* $x \in \times_{i \in N} X_i$ *is a composite equilibrium of* $\Gamma$ *under Nash equilibrium if* $\kappa(x, N) = |N|$.

**Proof:** It suffices to show that if all the strategies in $x$ are potentially dominant, $\{\rho_x(i)\}_{i \in N}$ is a partition of $N$.

First, observe that if $\kappa(x, N) = |N|$, no player is mapped to the empty set. Since each player must therefore be mapped onto some coalition that contains them, we have that $N \subseteq \bigcup_{S \in \{\rho_x(i)\}_{i \in N}} S$. Moreover, $\bigcup_{S \in \{\rho_x(i)\}_{i \in N}} S \subseteq N$ must hold as the codomain of $\rho_x$ is $2^N$. It follows that $\bigcup_{S \in \{\rho_x(i)\}_{i \in N}} S = N$.

Now suppose that $\{\rho_x(i)\}_{i \in N}$ fails to be a partition as there exist coalitions $S, T \in \{\rho_x(i)\}_{i \in N}$ such that $S$ and $T$ are not disjoint. Then, there exists a player $j$ such that $j \in S \cap T$. This implies that player $j$'s strategy in $x$ is part of the dominant strategies of both coalitions $S$ and $T$. This contradicts the requirement that $\rho_x$ be well-defined for all strategy profiles. ∎

*2.2 Notation and Definitions for Cooperative Games*

With the same player set as defined in the non-cooperative game, we can consider a cooperative game $(N, v)$ where $v$ is the worth function. We refer to the tuple $(S, \pi) \in 2^N \times \Pi(N)$ as an *embedded coalition* if $S \in \pi$. $v(S, \pi)$ is the value of the coalition $S$ given the partition $\pi$ and is defined only for embedded coalitions.

The worth function $v$ is *weakly superadditive* if for all pairs of embedded coalitions of the form $(S, \pi)$ and $(T, \pi)$ – that is, we require the coalitions $S$ and $T$ to be disjoint and belong to the same partition – the following holds: $v(S, \pi) + v(T, \pi) \leq v(S \cup T, \{S \cup T\} \cup \pi \setminus \{S, T\})$. $v$ is said to be *strictly superadditive* if the above always holds with strict inequality instead.

Given our assumption that players can form coalitions without incurring external costs, weak superadditivity is a minimal property to expect from worth functions, particularly in implementable cooperative games. For the underlying non-cooperative game, members of any given coalition can always play the same strategies as those played in a composite game prior to coalition formation but doing so jointly. Holding fixed the coalitional structure of the other players, this guarantees a value equal to the sum of their prior payoffs. In this regard, players are always weakly better off from forming coalitions.

When the coalitional structure of the other players does change, we may observe differences in what a coalition can feasibly achieve. $v$ is said to exhibit a *positive externality* if there exist disjoint embedded coalitions of the form $(R, \pi)$, $(S, \pi)$, and $(T, \pi)$ such that $v(R, \pi) < v(R, \{R\} \cup \{S \cup T\} \cup \pi \setminus \{R, S, T\})$. $v$ exhibits a *negative externality* if there exist disjoint embedded coalitions for which this strict inequality is reversed. Note that from this definition a worth function may exhibit both positive and negative externalities.

In general, intuition for these properties is best developed in the case of implementable cooperative games. For example, an externality in the worth function reflects how coalition formation/dissolution results in an aggregation/disaggregation of payoffs that incentivize a change in strategy, ultimately shifting the equilibrium to one that rewards different payoffs for all the players. Given this, we present a formal definition below of what it means for a cooperative game to be implementable with respect to normal-form games.

**Definition 4: Implementability**

Given a non-cooperative solution concept to be applied throughout, the cooperative game $(N, v)$ is *implementable under the solution concept* if there exists a normal-form game $(N, (X_i)_{i \in N}, (u_i)_{i \in N})$ such that the solution to each of its composite games yields payoffs to each embedded coalition equal to its value as defined in $(N, v)$. We say that $(N, (X_i)_{i \in N}, (u_i)_{i \in N})$ is *a (normal-form) implementation* of $(N, v)$ *under the non-cooperative solution concept*.

Suppose that $(N, (X_i)_{i \in N}, (u_i)_{i \in N})$ is an implementation of $(N, v)$ under Nash equilibrium such that all its composite equilibria consist only of strictly dominant strategies. Then, by definition, $v(S, \pi) = u_S(\sigma_S^*, \sigma_{\pi \setminus \{S\}}^*)$ must hold for all embedded coalitions $(S, \pi)$.

## 3. Example Implementations under Nash Equilibrium of Partition Function Form Games with $|N| = 2$ and $|N| = 3$

As we shall show, the task of constructing an implementation of a cooperative game under Nash equilibrium simplifies when we only consider possible candidates for which Lemmata 1 and 2 hold. When we assume such a normal-form game $\Gamma$, we can count the number of potentially dominant strategies in a strategy profile – that is, apply $\kappa$ – to determine whether or not it is a composite equilibrium of $\Gamma$. We can then define each player's payoff function accordingly such that payoffs to coalitions at composite equilibria coincide with their value in the cooperative game. Evidently, we must also ensure that these payoff functions defined at strategy profiles that are not composite equilibria do not contradict the requirements for Lemmata 1 and 2 to be applied. We hence proceed as follows.

For all players $i \in N$, we define their strategy set to be $X_i = \{\sigma_i^S | S \in 2^N, i \in S\}$. Critically, we assume that each coalition has a dominant strategy of the form $\sigma_S^* = \times_{i \in S} \sigma_i^S$. By extension, for any individual player, their dominant strategy is indexed by $\sigma_i^* = \sigma_i^i$. When this assumption holds, $\rho_x$ is well-defined for all strategy profiles. Thus, as long as $(u_i)_{i \in N}$ is defined such that $\sigma_S^* = \times_{i \in S} \sigma_i^S$ holds for each coalition and that $\sigma_S^*$ strictly dominates, we can invoke Lemmata 1 and 2.

It turns out that we can construct a parametric family of $(u_i)_{i \in N}$ that satisfies this requirement. Moreover, given a strictly superadditive partition function form game, $(N, v)$, each $(u_i)_{i \in N}$ in the family can be made to satisfy $v(S, \pi) = u_S(\sigma_S^*, \sigma_{\pi \setminus \{S\}}^*)$ for each embedded coalition.

For any given strategy profile $x$, let $u_i(x) = \theta f_i(x) + g_i(x)$ be the payoff function for each player $i$, where $\theta \in \mathbb{R}^+$ is a parameterizing constant. We define $f_i(x)$ and $g_i(x)$ by:

$$f_i(x) = \begin{cases} -|N| & \text{if } \rho_x(i) = \emptyset \\ |N| - \kappa(x, N) & \text{otherwise} \end{cases}$$

$$g_i(x) = \begin{cases} \frac{v(\rho_x(i), \{\rho_x(i)\}_{i \in N})}{|\rho_x(i)|} & \text{if } \kappa(x, N) = |N| \\ |\rho_x(i)| & \text{otherwise} \end{cases}$$

Moreover, we denote $f_S(x) = \sum_{i \in S} f_i(x)$ and $g_S(x) = \sum_{i \in S} g_i(x)$ for all coalitions $S \in 2^N$.

At first glance, the motivation behind such definitions may appear somewhat opaque. Intuitively, we require some control over the payoffs in the events where the player plays a potentially strictly dominant strategy and where all players play potentially strictly dominant strategies. This is achieved by $f_i(x)$ and $g_i(x)$ respectively, and their moderation by the parameter $\theta$ allows the possibility of payoffs $(u_i)_{i \in N}$ to support the condition that $\sigma_S^* = \times_{i \in S} \sigma_i^S$ holds for each coalition.

To illustrate this, let $\Gamma(N, v, \theta) = (N, (X_i)_{i \in N}, (u_i)_{i \in N})$ be the normal-form game in which the strategy sets and payoff functions are as defined before for some choice of worth function $v$ and parameter $\theta$. We demonstrate that there exist values of $\theta$ for which $\Gamma(N, v, \theta)$ implements $(N, v)$ under Nash equilibrium when $v$ is a strictly superadditive partition function for the cases where $|N| = 2$ and $|N| = 3$ – these being the smallest non-trivial cooperative game and the smallest non-trivial cooperative game that can exhibit externalities.

**Example 1 (Strictly Superadditive Partition Function Form Games for $|N| = 2$):**

Consider the following parameterization of the worth function for a two-player zero-normalized game:
$$v(\{1\}; \{\{1\}, \{2\}\}) = 0, v(\{2\}; \{\{1\}, \{2\}\}) = 0, v(\{1,2\}; \{\{1,2\}\}) = a, a > 0$$
Any two-player strictly superadditive partition function form game is uniquely identified by the parameter $a$. Below is the payoff matrix for $\Gamma(N, v, \theta)$ with $(N, v)$ as defined.

|  | $\sigma_2^{\{2\}}$ | $\sigma_2^{\{1,2\}}$ |
|---|---|---|
| $\sigma_1^{\{1\}}$ | 0, 0 | $\theta + 1, -2\theta$ |
| $\sigma_1^{\{1,2\}}$ | $-2\theta, \theta + 1$ | $\frac{a}{2}, \frac{a}{2}$ |

Table 1: Payoff Matrix for $\Gamma(N, v, \theta)$ when $|N| = 2$

Observe that when $\theta > max\left(\frac{a}{2} - 1, 1 - a\right)$, $\sigma_1^{\{1\}}$, $\sigma_2^{\{2\}}$, and $\sigma_1^{\{12\}} \times \sigma_2^{\{12\}}$ are indeed the strictly dominant strategies of player 1, player 2, and coalition $\{1,2\}$ respectively. Further inspection of the payoffs at composite equilibria shows that $\Gamma(N, v, \theta)$ implements $(N, v)$ under Nash equilibrium for all $\theta > max\left(\frac{a}{2} - 1, 1 - a\right)$.

**Example 2 (Strictly Superadditive Partition Function Form Games for $n = 3$):**

Now, consider the following parameterization of the worth function for a three-player zero-normalized game:
$$v(\{1\}; \{\{1\}, \{2\}, \{3\}\}) = 0, v(\{2\}; \{\{1\}, \{2\}, \{3\}\}) = 0, v(\{3\}; \{\{1\}, \{2\}, \{3\}\}) = 0$$
$$v(\{1,2\}; \{\{1,2\}, \{3\}\}) = a, v(\{3\}; \{\{1,2\}, \{3\}\}) = b, a > 0$$
$$v(\{1,3\}; \{\{1,3\}, \{2\}\}) = c, v(\{2\}; \{\{1,3\}, \{2\}\}) = d, c > 0$$
$$v(\{2,3\}; \{\{1\}, \{2,3\}\}) = e, v(\{1\}; \{\{1\}, \{2,3\}\}) = f, f > 0$$
$$v(\{1,2,3\}; \{\{1,2,3\}\}) = g, \quad g > max(0, a+b, c+d, e+f)$$

Any three-player strictly superadditive partition function form game is uniquely identified by the parameters $a, b, c, d, e, f$, and $g$. The payoff matrix for $\Gamma(N, v, \theta)$ with $(N, v)$ as defined is as follows:

| $\sigma_3^{\{3\}}$ | $\sigma_2^{\{2\}}$ | $\sigma_2^{\{1,2\}}$ | $\sigma_2^{\{2,3\}}$ | $\sigma_2^{\{1,2,3\}}$ |
|---|---|---|---|---|
| $\sigma_1^{\{1\}}$ | $0,0,0$ | $\theta+1,-3\theta,\theta+1$ | $\theta+1,-3\theta,\theta+1$ | $\theta+1,-3\theta,\theta+1$ |
| $\sigma_1^{\{1,2\}}$ | $-3\theta,\theta+1,\theta+1$ | $\frac{a}{2},\frac{a}{2},b$ | $-3\theta,-3\theta,2\theta+1$ | $-3\theta,-3\theta,2\theta+1$ |
| $\sigma_1^{\{1,3\}}$ | $-3\theta,\theta+1,\theta+1$ | $-3\theta,-3\theta,2\theta+1$ | $-3\theta,-3\theta,2\theta+1$ | $-3\theta,-3\theta,2\theta+1$ |
| $\sigma_1^{\{1,2,3\}}$ | $-3\theta,\theta+1,\theta+1$ | $-3\theta,-3\theta,2\theta+1$ | $-3\theta,-3\theta,2\theta+1$ | $-3\theta,-3\theta,2\theta+1$ |

| $\sigma_3^{\{1,3\}}$ | $\sigma_2^{\{2\}}$ | $\sigma_2^{\{1,2\}}$ | $\sigma_2^{\{2,3\}}$ | $\sigma_2^{\{1,2,3\}}$ |
|---|---|---|---|---|
| $\sigma_1^{\{1\}}$ | $\theta+1,\theta+1,-3\theta$ | $2\theta+1,-3\theta,-3\theta$ | $2\theta+1,-3\theta,-3\theta$ | $2\theta+1,-3\theta,-3\theta$ |
| $\sigma_1^{\{1,2\}}$ | $-3\theta,2\theta+1,-3\theta$ | $\theta+2,\theta+2,-3\theta$ | $-3\theta,-3\theta,-3\theta$ | $-3\theta,-3\theta,-3\theta$ |
| $\sigma_1^{\{1,3\}}$ | $\frac{c}{2},d,\frac{c}{2}$ | $\theta+2,-3\theta,\theta+2$ | $\theta+2,-3\theta,\theta+2$ | $\theta+2,-3\theta,\theta+2$ |
| $\sigma_1^{\{1,2,3\}}$ | $-3\theta,2\theta+1,-3\theta$ | $-3\theta,-3\theta,-3\theta$ | $-3\theta,-3\theta,-3\theta$ | $-3\theta,-3\theta,-3\theta$ |

| $\sigma_3^{\{2,3\}}$ | $\sigma_2^{\{2\}}$ | $\sigma_2^{\{1,2\}}$ | $\sigma_2^{\{2,3\}}$ | $\sigma_2^{\{1,2,3\}}$ |
|---|---|---|---|---|
| $\sigma_1^{\{1\}}$ | $\theta+1,\theta+1,-3\theta$ | $2\theta+1,-3\theta,-3\theta$ | $f,\frac{e}{2},\frac{e}{2}$ | $2\theta+1,-3\theta,-3\theta$ |
| $\sigma_1^{\{1,2\}}$ | $-3\theta,2\theta+1,-3\theta$ | $\theta+2,\theta+2,-3\theta$ | $-3\theta,\theta+2,\theta+2$ | $-3\theta,-3\theta,-3\theta$ |
| $\sigma_1^{\{1,3\}}$ | $-3\theta,2\theta+1,-3\theta$ | $-3\theta,-3\theta,-3\theta$ | $-3\theta,\theta+2,\theta+2$ | $-3\theta,-3\theta,-3\theta$ |
| $\sigma_1^{\{1,2,3\}}$ | $-3\theta,2\theta+1,-3\theta$ | $-3\theta,-3\theta,-3\theta$ | $-3\theta,\theta+2,\theta+2$ | $-3\theta,-3\theta,-3\theta$ |

| $\sigma_3^{\{1,2,3\}}$ | $\sigma_2^{\{2\}}$ | $\sigma_2^{\{1,2\}}$ | $\sigma_2^{\{2,3\}}$ | $\sigma_2^{\{1,2,3\}}$ |
|---|---|---|---|---|
| $\sigma_1^{\{1\}}$ | $\theta+1,\theta+1,-3\theta$ | $2\theta+1,-3\theta,-3\theta$ | $2\theta+1,-3\theta,-3\theta$ | $2\theta+1,-3\theta,-3\theta$ |
| $\sigma_1^{\{1,2\}}$ | $-3\theta,2\theta+1,-3\theta$ | $\theta+2,\theta+2,-3\theta$ | $-3\theta,-3\theta,-3\theta$ | $-3\theta,-3\theta,-3\theta$ |
| $\sigma_1^{\{1,3\}}$ | $-3\theta,2\theta+1,-3\theta$ | $-3\theta,-3\theta,-3\theta$ | $-3\theta,-3\theta,-3\theta$ | $-3\theta,-3\theta,-3\theta$ |
| $\sigma_1^{\{1,2,3\}}$ | $-3\theta,2\theta+1,-3\theta$ | $-3\theta,-3\theta,-3\theta$ | $-3\theta,-3\theta,-3\theta$ | $\frac{g}{3},\frac{g}{3},\frac{g}{3}$ |

Table 2: Payoff Matrix for $\Gamma(N,v,\theta)$ when $|N|=3$

Observe that $\sigma_1^{\{1\}}$, $\sigma_2^{\{2\}}$, $\sigma_3^{\{3\}}$, $\sigma_1^{\{1,2\}} \times \sigma_2^{\{1,2\}}$, $\sigma_1^{\{1,3\}} \times \sigma_3^{\{1,3\}}$, $\sigma_2^{\{2,3\}} \times \sigma_3^{\{2,3\}}$, and $\sigma_1^{\{1,2,3\}} \times \sigma_2^{\{1,2,3\}} \times \sigma_3^{\{1,2,3\}}$ are the strictly dominant strategies of players 1, 2, 3, and coalitions $\{1,2\}$, $\{1,3\}$, $\{2,3\}$, and $\{1,2,3\}$ respectively when the following holds:

$$\theta > \max\left(1, \frac{a}{2}-1, \frac{c}{2}-1, \frac{e}{2}-1, \frac{-b}{3}, \frac{-d}{3}, \frac{-f}{3}, \frac{g-3}{6}, \frac{1-a}{2}, \frac{1-c}{2}, \frac{1-e}{2}, \frac{a+2b-8}{4}, \frac{c+2d-8}{4}, \frac{e+2f-8}{4}, 4-g\right)$$

Again, by inspecting the payoffs at composite equilibria, we see that $\Gamma(N,v,\theta)$ implements $(N,v)$ under Nash equilibrium for these values of θ.

## 4. Theorems on the Existence of Implementations of Cooperative Games

**Theorem 1:** *Any strictly superadditive partition function form game $(N,v)$ is implementable under Nash equilibrium.*

**Proof:** Recall that for any given strictly superadditive partition function form game $(N,v)$, we define the normal-form game $\Gamma(N,v,\theta) = (N,(X_i)_{i\in N},(u_i)_{i\in N})$ whereby:

$$X_i = \{\sigma_i^S | S \in 2^N, i \in S\} \qquad (4)$$

$$u_i(x) = \theta f_i(x) + g_i(x) \tag{5}$$

$$f_i(x) = \begin{cases} -|N| & \text{if } \rho_x(i) = \emptyset \\ |N| - \kappa(x, N) & \text{otherwise} \end{cases} \tag{6}$$

$$g_i(x) = \begin{cases} \frac{v(\rho_x(i), \{\rho_x(i)\}_{i \in N})}{|\rho_x(i)|} & \text{if } \kappa(x, N) = |N| \\ |\rho_x(i)| & \text{otherwise} \end{cases} \tag{7}$$

Let us conjecture that there exists some $\theta \in \mathbb{R}^+$ such that for every coalition $S \in 2^N$, there further exists a dominant strategy of the form $\sigma_S^* = \times_{i \in S} \sigma_i^S$. Under this conjecture, $\rho_x$ is well-defined for all strategy profiles, and by extension, $\Gamma(N, v, \theta)$ is well-defined.

For any embedded coalition $(S, \pi)$, consider the strategy profiles $(\sigma_S^*, \sigma_{\pi \backslash \{S\}})$ and $(\sigma_S, \sigma_{\pi \backslash \{S\}})$, where $\sigma_S \neq \sigma_S^*$ and $\sigma_{\pi \backslash \{S\}}$ is fixed across both strategy profiles. For brevity, let $k = \kappa\big((\sigma_S^*, \sigma_{\pi \backslash \{S\}}), N\big)$, $l = \kappa\big((\sigma_S, \sigma_{\pi \backslash \{S\}}), N\big)$, and $m = \kappa\big((\sigma_S, \sigma_{\pi \backslash \{S\}}), S\big)$.

From (6), we have that $f_S(\sigma_S^*, \sigma_{\pi \backslash \{S\}}) = |S|(|N| - k)$ and $f_S(\sigma_S, \sigma_{\pi \backslash \{S\}}) = m(|N| - l) - (|S| - m)|N|$. Moreover, since $\kappa\big((\sigma_S^*, \sigma_{\pi \backslash \{S\}}), S\big) = |S|$, $\kappa\big((\sigma_S^*, \sigma_{\pi \backslash \{S\}}), \bigcup_{T \in \pi \backslash \{S\}} T\big) = k - |S| \leq \kappa\big((\sigma_S, \sigma_{\pi \backslash \{S\}}), \bigcup_{T \in \pi \backslash \{S\}} T\big)$. Applying (3), it follows that $k - |S| + m \leq l$. Altogether, $f_S(\sigma_S^*, \sigma_{\pi \backslash \{S\}}) - f_S(\sigma_S, \sigma_{\pi \backslash \{S\}}) \geq (|S| - m)(2|N| - k - m)$. As $|S| \geq m$, $|N| \geq k$, and $|N| > m$, it must be that $f_S(\sigma_S^*, \sigma_{\pi \backslash \{S\}}) - f_S(\sigma_S, \sigma_{\pi \backslash \{S\}}) \geq 0$.

Now, suppose $(\sigma_S^*, \sigma_{\pi \backslash \{S\}})$ and $(\sigma_S, \sigma_{\pi \backslash \{S\}})$ are such that $f_S(\sigma_S^*, \sigma_{\pi \backslash \{S\}}) = f_S(\sigma_S, \sigma_{\pi \backslash \{S\}})$. This holds if and only if $m = |S|$ and $k = l$, which in turn implies that $\{\rho_{(\sigma_S, \sigma_{\pi \backslash \{S\}})}(i)\}_{i \in N}$ is a strict refinement of $\{\rho_{(\sigma_S^*, \sigma_{\pi \backslash \{S\}})}(i)\}_{i \in N}$. Let us denote $\mathcal{P} = \{\rho_{(\sigma_S, \sigma_{\pi \backslash \{S\}})}(i)\}_{i \in S}$ to be the subset of $\{\rho_{(\sigma_S, \sigma_{\pi \backslash \{S\}})}(i)\}_{i \in N}$ for which $\bigcup_{T \in \mathcal{P}} T = S$. We consider the cases in which $k = |N|$ and $k < |N|$.

When $m = |S|$ and $k = l = |N|$, we obtain from (7) that $g_S(\sigma_S, \sigma_{\pi\{S\}}) = \sum_{T \in \mathcal{P}} v\big(T, \{\rho_{(\sigma_S, \sigma_{\pi\{S\}})}(i)\}_{i \in N}\big)$ and $g_S(\sigma_S^*, \sigma_{\pi \backslash \{S\}}) = v\big(S; \{\rho_{(\sigma_S^*, \sigma_{\pi \backslash \{S\}})}(i)\}_{i \in N}\big)$. By strict superadditivity, $\sum_{T \in \mathcal{P}} v\big(T, \{\rho_{(\sigma_S, \sigma_{\pi \backslash \{S\}})}(i)\}_{i \in N}\big) < v\big(S; \{\rho_{(\sigma_S^*, \sigma_{\pi \backslash \{S\}})}(i)\}_{i \in N}\big)$, so $g_S(\sigma_S, \sigma_{\pi \backslash \{S\}}) < g_S(\sigma_S^*, \sigma_{\pi \backslash \{S\}})$ and $u_S(\sigma_S, \sigma_{\pi \backslash \{S\}}) < u_S(\sigma_S^*, \sigma_{\pi \backslash \{S\}})$.

Alternatively, when $m = |S|$ and $k = l < |N|$, $g_S(\sigma_S^*, \sigma_{\pi \backslash \{S\}}) = |S|^2$ and $g_S(\sigma_S, \sigma_{\pi \backslash \{S\}}) = \sum_{T \in \mathcal{P}} |T|^2$. Since $0 < |T| < |S|$ for all $T \in \mathcal{P}$, it follows that $\sum_{T \in \mathcal{P}} |T|^2 < (\sum_{T \in \mathcal{P}} |T|)^2 = |S|^2$. Thus, $g_S(\sigma_S, \sigma_{\pi \backslash \{S\}}) < g_S(\sigma_S^*, \sigma_{\pi \backslash \{S\}})$ and once again $u_S(\sigma_S, \sigma_{\pi \backslash \{S\}}) < u_S(\sigma_S^*, \sigma_{\pi \backslash \{S\}})$.

Now, let us define the threshold value $\bar{\theta} \in \mathbb{R}^+$ by:

$$\bar{\theta} = \max\left(\left\{\left|\frac{g_S(\sigma_S, \sigma_{\pi \backslash \{S\}}) - g_S(\sigma_S^*, \sigma_{\pi \backslash \{S\}})}{f_S(\sigma_S^*, \sigma_{\pi \backslash \{S\}}) - f_S(\sigma_S, \sigma_{\pi \backslash \{S\}})}\right| S \in 2^N, \sigma_S \in X_S, \sigma_S \neq \sigma_S^*, \sigma_{\pi \backslash \{S\}} \in X_{\pi \backslash \{S\}}, f_S(\sigma_S^*, \sigma_{\pi \backslash \{S\}}) \neq f_S(\sigma_S, \sigma_{\pi \backslash \{S\}})\right\}\right) \tag{8}$$

Note that $\bar{\theta}$ exists as the maximum is taken over a finite set; the size of the set is bounded by the number of distinct pure strategy profiles in $\Gamma(N, v, \theta)$. So, when $(\sigma_S^*, \sigma_{\pi \backslash \{S\}})$ and $(\sigma_S, \sigma_{\pi \backslash \{S\}})$ are such that $f_S(\sigma_S^*, \sigma_{\pi \backslash \{S\}}) > f_S(\sigma_S, \sigma_{\pi \backslash \{S\}})$, $u_S(\sigma_S, \sigma_{\pi \backslash \{S\}}) < u_S(\sigma_S^*, \sigma_{\pi \backslash \{S\}})$ as long as $\theta > \bar{\theta}$. At these values of $\theta$, the conjecture for $\Gamma(N, v, \theta)$ holds.

In fact, for strict superadditivity, we have demonstrated that the dominant strategy of each coalition strictly dominates. Since $h(x) = \{\rho_x(i)\}_{i \in N}$ is a bijection between the sets $\{x|\kappa(x,N) = |N|\}$ and $\Pi(N)$, it follows from Lemmata 1 and 2 that there exists a bijection between the set of partitions and the set of composite equilibria of $\Gamma(N, v, \theta)$. Thus, for any embedded coalition $(S, \pi)$, there exists a unique composite equilibrium of $\Gamma(N, v, \theta)$, $x$, such that $\{\rho_x(i)\}_{i \in N} = \pi$. Since $\kappa(x,N) = |N|$, we have that $u_S(x) = \sum_{i \in S} \frac{v(\rho_x(i), \{\rho_x(i)\}_{i \in N})}{|\rho_x(i)|} = v(S, \pi)$. We conclude that $\Gamma(N, v, \theta)$ is an implementation of $(N, v)$ under Nash equilibrium for $\theta > \bar{\theta}$. ∎

The construction of $\Gamma(N, v, \theta)$ by (4) - (7) given any cooperative game $(N, v)$ and $\theta > \bar{\theta}$, for $\bar{\theta}$ defined in (8), can in fact be used to prove existence of implementations for other classes of cooperative games under other non-cooperative solution concepts as the following theorems demonstrate.

**Theorem 2:** *Any weakly superadditive characteristic function form game $(N, v)$ is implementable under Nash equilibrium.*

**Proof:** Observe that for any coalition $S$ in $\Gamma(N, v, \theta)$, $\sigma_S^*$ is no longer guaranteed to be a strictly dominant strategy since weak superadditivity of the characteristic function implies that there may exist distinct strategy profiles $(\sigma_S^*, \sigma_{\pi \setminus \{S\}})$ and $(\sigma_S, \sigma_{\pi \setminus \{S\}})$ such that $f_S(\sigma_S^*, \sigma_{\pi \setminus \{S\}}) = f_S(\sigma_S, \sigma_{\pi \setminus \{S\}})$ and $g_S(\sigma_S, \sigma_{\pi \setminus \{S\}}) = \sum_{T \in \mathcal{P}} v\left(T, \{\rho_{(\sigma_S, \sigma_{\pi \setminus \{S\}})}(i)\}_{i \in N}\right) = v\left(S; \{\rho_{(\sigma_S^*, \sigma_{\pi \setminus \{S\}})}(i)\}_{i \in N}\right) = g_S(\sigma_S^*, \sigma_{\pi \setminus \{S\}})$. By construction, this is the only case in which $u_S(\sigma_S, \sigma_{\pi \setminus \{S\}}) < u_S(\sigma_S^*, \sigma_{\pi \setminus \{S\}})$ may not hold.

Here, $(\sigma_S^*, \sigma_{\pi \setminus \{S\}})$ and $(\sigma_S, \sigma_{\pi \setminus \{S\}})$ are both Nash equilibria in the same composite game $\Gamma(N, v, \theta)_{\{\rho_{(\sigma_S^*, \sigma_{\pi \setminus \{S\}})}(i)\}_{i \in N}}$. Moreover, it must be true that $\kappa\left((\sigma_S^*, \sigma_{\pi \setminus \{S\}}), N\right) = \kappa\left((\sigma_S, \sigma_{\pi \setminus \{S\}}), N\right) = |N|$. It follows that for any coalition $U \in \{\rho_{(\sigma_S^*, \sigma_{\pi \setminus \{S\}})}(i)\}_{i \in N \setminus S} = \{\rho_{(\sigma_S, \sigma_{\pi \setminus \{S\}})}(i)\}_{i \in N \setminus S}$, $f_U(\sigma_S^*, \sigma_{\pi \setminus \{S\}}) = f_U(\sigma_S, \sigma_{\pi \setminus \{S\}})$ and $g_U(\sigma_S^*, \sigma_{\pi \setminus \{S\}}) = v\left(U; \{\rho_{(\sigma_S^*, \sigma_{\pi \setminus \{S\}})}(i)\}_{i \in N}\right) = v(U) = v\left(U; \{\rho_{(\sigma_S, \sigma_{\pi \setminus \{S\}})}(i)\}_{i \in N}\right) = g_U(\sigma_S, \sigma_{\pi \setminus \{S\}})$ as characteristic functions do not exhibit externalities. Hence, payoffs to coalitions are identical across both Nash equilibria and remain equal to the values as determined by the worth function. ∎

**Theorem 3:** *Any strictly superadditive partition function form game $(N, v)$ is implementable under rationalizability.*

**Proof:** Observe that under the conditions noted in Lemmata 1 and 2, an immediate corollary is that a strategy profile $x \in \times_{i \in N} X_i$ is a composite equilibrium of $\Gamma$ under rationalizability if and only if $\kappa(x, N) = |N|$. So, Theorem 3 follows directly from the proof of Theorem 1. ∎

**Theorem 4:** *Any weakly superadditive partition function form game $(N, v)$ is implementable under trembling-hand perfect equilibrium.*

**Proof:** As detailed in the proof of Theorem 2, we are no longer guaranteed that each coalition has a strictly dominant strategy when the worth function is weakly superadditive. However, it remains that each coalition has a dominant strategy, which generate a unique trembling-hand perfect equilibrium in each composite game. It is easy to check that payoffs at these composite equilibria coincide with the coalitions' values as defined by $v$. ∎

# 5. Discussion of the Selection of Non-Cooperative Solution Concept for the Implementation of a Cooperative Game

Now that implementability has been established for different classes of cooperative games under various non-cooperative solution concepts, we turn our attention towards the appropriate choice of solution concept. The same non-cooperative game may implement several distinct cooperative games depending on the solution concept applied, which in turn leads to even more cooperative solutions to consider. Pinning down the most appropriate worth function that models the economic interaction at hand is thus a prerequisite to filtering out inadequate cooperative solutions. For this, we will need to pay greater attention to the non-cooperative solution concept adopted.

Much of cooperative game theory has historically focused on $\alpha$-characteristic function form games, the class of cooperative games implementable under minimax outcomes. While such games offer a viable means of summarizing zero-sum interactions, we argue that they produce questionable results for general-sum games. For general-sum settings, we instead recommend cooperative games implementable under Nash equilibrium as a more sensible choice of model. We illustrate these arguments through a classic treatment of Cournot and Bertrand competition.[8] For both minimax outcomes and Nash equilibrium, we solve for the respective worth functions and compare the cooperative solutions that follow.

So, let $N = \{1, \ldots, n\}$ be a set of firms producing an identical good with each firm $i \in N$ paying a constant marginal cost of production $c_i$ such that $c_{i-1} < c_i < c_{i+1}$. Firms face a bounded inverse demand curve given by

$$p(Q) = \begin{cases} a - bQ & \text{if } 0 \leq Q \leq \frac{a}{b} \\ 0 & \text{if } Q > \frac{a}{b} \end{cases} \quad (9)$$

with $Q = \sum_{i=1}^{n} q_i$ denoting the total quantity of the good supplied to the market. We additionally require that $a > c_n$ for an interior solution. Costless collusion means that firms can choose to cooperate with other firms and form cartels that act as a single profit-maximizing entity. Within each cartel, optimal behavior implies that only the lowest-cost firm supplies a positive quantity of the good; other firms in the cartel produce nothing. The resulting profits are expected to be redistributed across members, else there would be little incentive for firms to cooperate in the first place.

Traditional questions in cooperative game theory ask what this allocation should look like. Is it stable in the sense that no firm would benefit from a deviation from the allocation? And is it fair insofar as it satisfies requirements that we may deem desirable from a normative perspective? As will be shown, leading solution concepts in cooperative game theory such as the core and the Shapley value make little sense as potentially stable and fair outcomes when the implementation is under maximin outcomes.

### 5.1 Oligopolistic Competition under Maximin Outcomes

In both Cournot and Bertrand competition, any cartel can threaten to reduce the profits of the other firms to zero either by flooding the market with goods to drive the price down or by undercutting all their competitors to attract complete market share. Such threats are hardly credible, and yet the cooperative game implemented under maximin outcomes treats firms as taking them seriously. In particular, the worth function is a characteristic function given by:

$$v(S) = \begin{cases} \frac{(a-c_1)^2}{4b} & if\ S = N \\ 0 & otherwise \end{cases} \quad (10)$$

---

[8] More recent treatment of oligopolistic competition from a cooperative lens includes Abe (2021), Chander (2020), Grabisch and Funaki (2012), and Kohlberg and Nehman (2021).

Notwithstanding the behavior of the other firms, the best that any cartel other than the grand cartel can guarantee itself here is zero. Only the grand cartel is unexposed to threats and thus guarantees itself monopoly profits. As noted in Ray (2007), such a worth function delivers strange cooperative solutions with little economic interpretation. For example, by the symmetry of (10), the Shapley value of this game allocates each firm an equal share of the monopoly profit, despite heterogeneity in production costs across firms. Perhaps more striking is that the core admits any non-negative imputation, so even one that allocates the minimal surplus to the lowest-cost producer will supposedly not face rejection.

### 5.2 Cournot Competition Under Nash Equilibrium

Meanwhile, under Nash equilibrium, the worth functions for Cournot and Bertrand competition are distinct as the profits of each cartel in any given composite game are distinct across the two types of competition. This results in cooperative solutions that differ across competitive environments, reflecting the variations in market power across firms.

For any embedded cartel $(S, \pi)$, let its cost of production be denoted $c_S = c_{min(S)}$, which we recall is the cost of production of the most efficient firm in the cartel. To ensure the existence of an interior solution in which each cartel of any given composite game produces a positive quantity, let us assume that for all embedded cartels:

$$\frac{a + \sum_{T \in \pi \setminus \{S\}} c_T}{|\pi|} > c_S \quad (11)$$

For example, (11) can be satisfied by a sufficiently high demand $a$. The worth function for Cournot competition under Nash equilibrium is given by:[9]

$$v(S, \pi) = \frac{(a + \sum_{T \in \pi \setminus \{S\}} c_T - |\pi| c_S)^2}{(|\pi| + 1)^2 b} \quad (12)$$

**Proposition 1:** *At the interior solution, $v(S, \pi)$ as defined in (12) is a partition function and the only externalities that it exhibits are positive.*

**Proof:** For any two embedded cartels $(S, \pi)$ and $(S, \pi')$, where $\pi'$ is a strict refinement of $\pi$, we wish to show that $v(S, \pi) > v(S, \pi')$. Assume not. Then it follows that:

$$\frac{a + \sum_{T \in \pi \setminus \{S\}} c_T - |\pi| c_S}{|\pi| + 1} \leq \frac{a + \sum_{T \in \pi' \setminus \{S\}} c_T - |\pi'| c_S}{|\pi'| + 1} \quad (13)$$

Rearranging (13) and simplifying, we get:

$$(|\pi'| - |\pi|)a + (|\pi'| + 1)\sum_{T \in \pi} c_T \leq (|\pi| + 1)\sum_{T \in \pi'} c_T \quad (14)$$

Observe that $\{c_T | T \in \pi\} \subset \{c_T | T \in \pi'\}$ and let $\mathcal{C} = \{c_T | T \in \pi\} \setminus \{c_T | T \in \pi'\}$. We can therefore express (14) as:

$$(|\pi'| - |\pi|)(a + \sum_{T \in \pi} c_T) \leq (|\pi| + 1)\sum_{C_U \in \mathcal{C}} c_U \quad (15)$$

---

[9] A full derivation is provided in Appendix A.

Now, note that it is always possible to construct a partition $\hat{\pi}$ such that $|\hat{\pi}| = |\pi| + 1$ and $\{c_T | T \in \hat{\pi}\} = \{c_T | T \in \pi\} \cup \{c_U\}$ for any $c_U \in \mathcal{C}$. Thus, by (11), for any $c_U \in \mathcal{C}$:

$$\frac{a + \sum_{T \in \pi} c_T}{|\hat{\pi}|} > c_U \qquad (16)$$

Rearranging (16) and summing across all $c_U \in \mathcal{C}$, we get $(|\pi| + 1) \sum_{c_U \in \mathcal{C}} c_U < (|\pi'| - |\pi|)(a + \sum_{T \in \pi} c_T)$, a contradiction. ∎

Although we may expect that firms would not favor the collusion of their competitors lest they consolidate greater market power, we should note that there are no economies of scale in this model, and so no combination of high-cost firms can overtake the competitiveness of the lower-cost producers. Instead, collusion in this example is effectively an agreement for productively inefficient firms within a cartel to cease production, the effects of which spill over to the entire market as a reduction in competition. Since all firms may face this positive externality, we should additionally expect that collusion here is difficult to maintain. That is, regarding cooperative solutions, we should expect the core to be small or indeed empty.

But for partition function form games, whether the grand coalition is stable depends on how players expect other players to behave following a deviation. If firms believe that their deviation from the grand cartel will not trigger further deviations, then they may proceed as planned. But if firms anticipate a complete dissolution of the grand cartel following their deviation, then they may refrain from doing so. Two classic solution concepts based on the core immediately follow. Although much discussion is to be had regarding the legitimacy of such expectations and the suitability of the refinements, we present them here as examples of cooperative solutions that are more plausible predictions of collusive behavior than those discussed previously in the minimax case.[10]

**Definition 5: $\gamma$-Core (Hart and Kurz, 1983)**

For any partition function form game $(N, v)$, an imputation $y = (y_1, \ldots, y_n)$ lies in the $\gamma$-core if for all coalitions $S \in 2^N$, $\sum_{i \in S} y_i \geq v(S, \{S\} \cup [N \backslash S])$.

**Definition 6: $\delta$-Core (Hart and Kurz, 1983)**

For any partition function form game $(N, v)$, an imputation $y = (y_1, \ldots, y_n)$ lies in the $\delta$-core if for all coalitions $S \in 2^N$, $\sum_{i \in S} y_i \geq v(S, \{S\} \cup \{N \backslash S\})$.

**Proposition 2:** *For the cooperative game $(N, v)$ with $v$ defined by (12), the $\gamma$-core is always non-empty.*

**Proof:** See Appendix B.

**Proposition 3:** *For the cooperative game $(N, v)$ with $v$ defined by (12), the $\delta$-core is empty if any of the following hold: $|N|$ is asymptotically large, $a$ is asymptotically large for $|N| > 2$, or the costs of production for each firm $\{c_i\}_{i \in N}$ are asymptotically close together for $|N| > 2$.*

By Definition 6, the $\delta$-core will be empty if the following holds:

---

[10] For an axiomatic treatment of how coalitions may form expectations, see Bloch and van den Nouweland (2014). From their proposed set of axioms, they highlight the $\delta$-core (or the projection core as presented in the paper) as being a natural refinement of the core for partition function form games. Other refinements of the core can be found in Huang and Sjöström (2003) and Hafalir (2007). For refinements inspired by the Shapley value, see Myerson (1977), Owen (1977), Bolger (1989), and more recently McQuillin (2009).

$$\sum_{i \in N} v(\{i\}, \{\{i\}\} \cup \{N \setminus \{i\}\}) = \frac{(a+c_2-c_1)^2}{9b} + \sum_{i=2}^{n} \frac{(a+c_1-c_i)^2}{9b} > \frac{(a-c_1)^2}{4b} \qquad (17)$$

Since $a > c_i$ for all firms $i \in N$, the left-hand side is a monotonically increasing and divergent sum in $n$. Thus, there exists $\tilde{n}$ such that for all $n \geq \tilde{n}$, (17) holds. Likewise, as $a \to \infty$, $\frac{1}{a^2}\left(\frac{(a+c_2-c_1)^2}{9b} + \sum_{i=2}^{n} \frac{(a+c_1-c_i)^2}{9b} - \frac{(a-c_1)^2}{4b}\right) \to \frac{n}{9b} - \frac{1}{4b} > 0$ for $n > 2$. Finally, let $c_k = c_1 + \sum_{i=1}^{k-1} \epsilon_i$ for all $k \in \{2, \dots, n\}$. As $\epsilon_i \to 0$ for all $i \in \{1, \dots, n-1\}$, $\frac{(a+c_2-c_1)^2}{9b} + \sum_{i=2}^{n} \frac{(a+c_1-c_i)^2}{9b} \to \frac{na^2}{9b} > \frac{(a-c_1)^2}{4b}$ for $n > 2$. ∎

Hence, stability of the grand cartel is possible for any number of firms and any level of demand and production costs satisfying (11) when firms expects the grand cartel to dissolve fully following a deviation. But when firms expect no reaction from a deviation, then the stability of the grand cartel would require that the number of firms be small, market demand be relatively low, or there be significant heterogeneity in marginal costs. Otherwise, the profit incentive for firms from the positive externality is too high to support total collusion.

*5.3 Bertrand Competition Under Nash Equilibrium*

For Bertrand competition under Nash equilibrium, we suppose that a firm will choose not to produce when the price is exactly equal to their cost of production. Thus, only the lowest-cost producer will provide positive quantities to the market, selling at the second-lowest marginal cost of production. By the assumption made in (11), this is always below the monopoly price. The worth function is as follows:

$$v(S, \pi) = \begin{cases} \frac{(a-c_1)^2}{4b} & \text{if } S = N \\ (\min(\{c_T | T \in \pi \setminus \{S\}\}) - c_S)\left(\frac{a - \min(\{c_T | T \in \pi \setminus \{S\}\})}{b}\right) & \text{if } c_S = c_1 \\ 0 & \text{otherwise} \end{cases} \qquad (18)$$

This worth function is a characteristic function – just like the worth function implemented under maximin outcomes. But what causes the lack of externalities here is not the existence of incredible threats but rather the fact that no amount of collusion between inefficient firms can create a cartel that can compete with the lowest-cost producer and earn abnormal profits. Again, this is a result that follows from the absence of economies of scale in this model. In this case, coalition formation reduces market competition only when the lowest-cost and second lowest-cost cartels collude to raise market prices. Given all this, we should expect cooperative solutions to favor efficient firms as the following propositions demonstrate.

**Proposition 4:** *Let $(N, v)$ be the cooperative game with $v$ defined by (18). Its Shapley value, denoted $\phi(N, v)$, allocates greater profits to lower-cost firms; that is, $\phi_i(N, v) > \phi_j(N, v)$ for all $i, j \in N$ such that $i < j$.*

**Proof:** Recall that the Shapley value for any player $i \in N$ in a characteristic function form game $(N, v)$ as given by the random ordering approach is:

$$\phi_i(N, v) = \sum_{S \subseteq N \setminus \{i\}} \frac{|S|!(n-|S|-1)!}{n!} (v(S \cup \{i\}) - v(S))$$

For any firm $i, j \in N$ such that $i < j$, observe that for all cartels $S \in N \setminus \{i, j\}$, $v(S \cup \{i\}) \geq v(S \cup \{j\})$. Thus, $v(S \cup \{i\}) - v(S) \geq v(S \cup \{j\}) - v(S)$ and $v(S \cup \{i, j\}) - v(S \cup \{i\}) \leq v(S \cup \{i, j\}) - v(S \cup \{j\})$, holding strictly in the case where $\min(\{c_k | k \in N \setminus S\}) = c_i$.

It follows immediately that $\phi_i(N, v) > \phi_j(N, v)$. ∎

**Proposition 5:** *Let $(N, v)$ be the cooperative game with $v$ defined by (18). Its Shapley value, $\phi(N, v)$, is stable insofar as it lies in the core of $(N, v)$.*

**Proof:** By Shapley's theorem (1971), it suffices to show that $(N, v)$ is a convex game; that is, $v(S) + v(T) \leq v(S \cup T) + v(S \cap T)$ for all cartels $S, T \subseteq N$.

As a convention, let $v(\emptyset) = 0$. For any two cartels $S, T \subseteq N$, where $S$ and $T$ need not be disjoint, denote $i$ and $j$ as the lowest-cost producers outside cartels $S$ and $T$ respectively, where $i$ and $j$ similarly might not be distinct. By De Morgan's law, $\{i, j\} \subseteq N \backslash (S \cap T)$. Thus, $v(S \cap T) = \min(v(S), v(T))$. Without loss of generality, let $v(S \cap T) = v(S)$. Since $v(T) \leq v(S \cup T)$, it follows that $v(S) + v(T) \leq v(S \cup T) + v(S \cap T)$. ∎

## 6. Conclusion

This paper synthesizes previous literature on the determination of non-cooperative foundations for cooperative games and provides a generalized framework that extends the classical implementation of characteristic function form games under maximin behavior to a larger class of cooperative games under various other non-cooperative solution concepts. Motivation for this extension is then demonstrated through the example of oligopolistic competition in which cooperative solutions to a game implemented under Nash equilibrium are shown to offer more reasonable predictions of cartel formation than those derived under maximin outcomes.

We hope that these contributions help to inspire greater interest and encouragement for the practice of cooperative game theory. Mainstream economics research and teaching arguably pay relatively little attention to cooperative game theory compared to its non-cooperative counterpart despite the intimate connections between the two disciplines as emphasized in this paper. Part of this lack of recognition may be attributable to the failure of characteristic function form games as the most established model in cooperative game theory to accommodate externalities that arise when agreements to cooperate are formed or broken (Maskin, 2016). We hope that our inclusion and focus on partition function form games will help to address such concerns when it comes to cooperative games more broadly.

### Appendix A: Derivation of the Worth Function for Cournot Competition Under Nash Equilibrium

Consider the composite game played by partition $\pi \in \Pi(N)$. From (9), the objective function of any given cartel in $S \in \pi$ is given by:

$$\Pi(q_S) = \max_{q_S \geq 0} \{0, (a - bQ)q_S - c_S q_S\} \quad (19)$$

The interior solution to (19) satisfies the first order condition for each cartel:

$$\frac{a - c_S}{b} = 2q_S + \sum_{T \in \pi \backslash \{S\}} q_T \quad (20)$$

Summing across (20) for all cartels $S \in \pi$, we get:

$$\frac{|\pi|a - \sum_{S \in \pi} c_S}{(|\pi| + 1)b} = \sum_{S \in \pi} q_S \quad (21)$$

Subtracting (21) from (20) for some choice of $S \in \pi$, we have that:

$$q_S = \frac{a + \sum_{T \in \pi \backslash \{S\}} c_T - |\pi| c_S}{(|\pi| + 1)b}$$

Under assumption (11) for the existence of an interior solution, total market quantity and market prices are given by:

$$Q = \frac{|\pi|a - \sum_{S \in \pi} c_S}{(|\pi|+1)b} \qquad p = \frac{a + \sum_{S \in \pi} c_S}{|\pi|+1}$$

Thus, the value of any given embedded cartel, or equivalently their profits earned, is given by (12): $v(S, \pi) = \frac{(a + \sum_{T \in \pi \setminus \{S\}} c_T - |\pi| c_S)^2}{(|\pi|+1)^2 b}$.

## Appendix B: Proof of Proposition 2

Observe that the γ-core is equivalent to the core of the γ-characteristic function form game derived from the same implementation, which we shall denote $(N, v^\gamma)$. For all cartels $S$:

$$v^\gamma(S) = \frac{(e^S \cdot c)^2}{(|N \setminus S| + 2)^2 b}$$

where $e^S, c \in \mathbb{R}^{n+1}$ and $e_i^S = \begin{cases} 1 & \text{if } i \in N \setminus S \text{ or } i = n+1 \\ -|N \setminus S| - 1 & \text{if } i = \min(S) \text{ for } S \neq \emptyset, e^\emptyset = 0_{n+1} \text{ is the} \\ 0 & \text{otherwise} \end{cases}$

zero vector, and $c = (c_1, \ldots, c_n, a)$. Now, define $e^{i,j} \in \mathbb{R}^{n+1}$ where $e_k^{i,j} = \begin{cases} -1 & \text{if } k = i \\ 1 & \text{if } k = j \\ 0 & \text{otherwise} \end{cases}$.

We let $E$ be the set of vectors such that $E = \{e^{i,j} | i < j\} \cup \{e^S | S \in 2^N\}$. We denote the vector of all ones in $\mathbb{R}^{n+1}$ by $1_{n+1}$. Observe that for all $e \in E$, $e \cdot 1_{n+1} = 0$ and $e \cdot c \geq 0$, the latter inequality following from the ordering of marginal costs and assumption (11). Let $cone(E)$ denote the convex cone of vectors in $E$. Clearly, for all $e \in cone(E)$, $e \cdot 1_{n+1} = 0$ and $e \cdot c \geq 0$ must hold as well. Moreover, let $A$ be a $(n+1) \times \left(\binom{n}{2} + 2^n\right)$ matrix where each column vector $A_j$ is a distinct element of the set $E$.

**Lemma 3:** For any vector $w \in \mathbb{R}^{n+1}$ such that $A^T w \leq 0$, then $w$ is such that: **(a)** $w_i \geq w_j$ for all $i < j$ and **(b)** $w_i \geq \frac{\sum_{j \in C} w_j}{|C|}$ for all $i \neq n+1$ where $C \subseteq \{1, \ldots, n+1\}$ and $n+1 \in C$.

**Proof:** Take some $w$ such that $A^T w \leq 0$. As $e^{i,j} \cdot w \leq 0$ for all $i < j$, it follows that $w_i \geq w_j$.

Now, for any $C \subseteq \{1, \ldots, n+1\}$ such that $n+1 \in C$, choose some index $i \neq n+1$. Let us define $\tilde{C} = \{j | j < i, j \in N \setminus C\}$. Observe that $w_i \geq \frac{\sum_{j \in C \cup \tilde{C}} w_j}{|C \cup \tilde{C}|}$ as $e^S \cdot w \leq 0$ for $S = \{i\} \cup N \setminus (C \cup \tilde{C})$. But since $w_j \geq w_i$ for all $j \in \tilde{C}$, it follows that $w_i \geq \frac{\sum_{j \in C} w_j}{|C|}$. ∎

Lemma 3 implies that if $A^T w \leq 0$, then any entry in $w$ other than the last is weakly greater than the mean of any set of entries in $w$, provided that this set includes the last entry.

**Lemma 4:** For any vector $b \in \mathbb{R}^{n+1}$ such that $b \cdot 1_{n+1} = 0$ and $b_{n+1} = \max_{i \in \{1, \ldots, n+1\}} b_i$, $b \in cone(E)$.

**Proof:** Assume not. By Farkas' lemma, there exists a vector $w \in \mathbb{R}^{n+1}$ such that $\langle b, w \rangle > 0$ and $A^T w \leq 0$. Let us define $b^+, b^- \in \mathbb{R}^{n+1}$ by $b_i^+ = \begin{cases} b_i & \text{if } b_i > 0 \\ 0 & \text{otherwise} \end{cases}$ and $b_i^- = \begin{cases} b_i & \text{if } b_i < 0 \\ 0 & \text{otherwise} \end{cases}$.
We denote the sum of their entries by $B = \sum_i b_i^+ = -\sum_i b_i^-$.

Now, let us define $\overline{b^+}$ and $\underline{b^-}$ by the following construction. First, initialize both vectors to be the zero vector. Then, for $\overline{b^+}$, set $\overline{b^+_{n+1}} = b_{n+1}$, and then iterating from $i = 1$ onwards, check if $b_i^+$ is non-zero. If so and $\sum_{j=1}^{i-1} \overline{b_j^+} + 2b_{n+1} \leq B$ holds true, set $\overline{b_i^+}$ equal to $b_{n+1}$. Continue until the first index $l$ such that $b_l^+$ is non-zero and $\sum_{j=1}^{l-1} \overline{b_j^+} + 2b_{n+1} > B$. At that point, set $\overline{b_l^+}$ equal to $B(mod\ b_{n+1})$. Meanwhile for $\underline{b^-}$, let $m$ be the last index such that $b_m^-$ is non-zero. Set $\underline{b_m^-}$ equal to $-B$. Observe that both these constructions are made possible under the requirements that $b \cdot 1_{n+1} = 0$ and $b_{n+1} = max_{i \in \{1,\ldots,n+1\}} b_i$.

From Lemma 3a, we note that $\langle b, w \rangle = \langle b^+, w \rangle + \langle b^-, w \rangle \leq \langle \overline{b^+}, w \rangle + \langle \underline{b^-}, w \rangle$.

Now, suppose that $l < m$. Let $d \in \mathbb{R}^{n+1}$ where $d_i = \begin{cases} -(b_{n+1} - B(mod\ b_{n+1})) & if\ i = l \\ b_{n+1} - B(mod\ b_{n+1}) & if\ i = m \\ 0 & otherwise \end{cases}$.

By Lemma 3b, $\langle \overline{b^+} + \underline{b^-} - d, w \rangle \leq 0$ and $\langle d, w \rangle \leq 0$. Thus, $\langle b, w \rangle \leq 0$, a contradiction.

Meanwhile, when $l > m$, let us define $d' \in \mathbb{R}^{n+1}$ where $d'_i = \begin{cases} B(mod\ b_{n+1}) & if\ i = l \\ -B(mod\ b_{n+1}) & if\ i = m \\ 0 & otherwise \end{cases}$.

By Lemma 3b, $\langle \overline{b^+} + \underline{b^-} - d', w \rangle \leq 0$ and $\langle d', w \rangle \leq 0$. Thus, $\langle b, w \rangle \leq 0$, a contradiction. ∎

**Lemma 5:** For all cartels $S, T \in 2^N$, $\left( \frac{e^{S \cup T}}{|N \setminus (S \cup T)|+2} + \frac{e^{S \cap T}}{|N \setminus (S \cap T)|+2} \right) - \left( \frac{e^S}{|N \setminus S|+2} + \frac{e^T}{|N \setminus T|+2} \right) \in Cone(E)$.

**Proof:** For ease of notation, let $A = N \setminus S$, $B = N \setminus T$ and $b = \left( \frac{e^{S \cup T}}{|A \cap B|+2} + \frac{e^{S \cap T}}{|A \cup B|+2} \right) - \left( \frac{e^S}{|A|+2} + \frac{e^T}{|B|+2} \right)$. Now, $b_{n+1}$ is non-negative as $\frac{1}{|A|+2} + \frac{1}{|B|+2} \leq \frac{1}{|A \cap B|+2} + \frac{1}{|A \cup B|+2}$, which follows from the results that $|A| + |B| = |A \cap B| + |A \cup B|$ and $|A||B| \geq |A \cap B||A \cup B|$ for any $A, B \in 2^N$. We proceed to show that $b_i \leq b_{n+1}$ for all $i \in N$.

Observe that by construction, for any index $i \in N \setminus (S \cup T)$, $b_i = b_{n+1}$.

Now, consider the indices $i \in S$. Let $i_S = min(S)$ be the index of the negative entry in $e^S$. Either $i_{S \cup T} = i_S$ or $i_{S \cup T} = i_T$ or both. Without loss of generality, let us assume $i_{S \cup T} = i_S$ such that $i_{S \cup T} = i_S \leq i_T \leq i_{S \cap T}$.

Suppose $i_S = i_T$. By extension, $i_S = i_{S \cap T}$. It follows that $b_{i_S} = \left( \frac{-(|A \cap B|+1)}{|A \cap B|+2} + \frac{-(|A \cup B|+1)}{|A \cup B|+2} \right) - \left( \frac{-(|A|+1)}{|A|+2} + \frac{-(|B|+1)}{|B|+2} \right) = b_{n+1}$. Alternatively, suppose that $i_S \notin T$. Then, it follows that $b_{i_S} = \left( \frac{-(|A \cap B|+1)}{|A \cap B|+2} + \frac{1}{|A \cup B|+2} \right) - \left( \frac{-(|A|+1)}{|A|+2} + \frac{1}{|B|+2} \right) = b_{n+1}$.

Otherwise, for indices $i \in S \setminus \{i_S\}$, if $i \notin T$, then $i \neq S \cap T$, so $b_i = \frac{1}{|A \cup B|+2} - \frac{1}{|B|+2} \leq 0$ as $|A \cup B| \geq |B|$. If $i \in T$ and $i = i_T$, then $i = i_{S \cap T}$ and $b_i = \frac{-(|A \cup B|+1)}{|A \cup B|+2} - \frac{-(|B|+1)}{|B|+2} = \frac{1}{|A \cup B|+2} - \frac{1}{|B|+2} \leq 0$. Meanwhile, if $i \in T$ and $i \neq i_T$, then $i \in S \cup T$ and $i \in S \cap T$, and $b_i \leq 0$.

Finally, consider the indices $i \in T \setminus S$. If $i \neq i_T$, then $b_i = \frac{1}{|A \cup B|+2} - \frac{1}{|A|+2} \leq 0$ as $|A \cup B| \geq |A|$. Likewise, if $i = i_T$, then $b_i = \frac{1}{|A \cup B|+2} - \left( \frac{1}{|A|+2} + \frac{-(|B|+1)}{|B|+2} \right) \leq 0$.

Hence, $b_{n+1} = max_{i \in \{1,\ldots,n+1\}} b_i$. Furthermore, since $b$ is a linear combination of elements in $E$, $b \cdot 1_{n+1} = 0$. Applying Lemma 4, we conclude that $b \in Cone(E)$. ∎

Since $e \cdot c \geq 0$ for all $e \in cone(E)$, from Lemma 5, it follows that for all cartels $S, T \subseteq N$,

$$\sqrt{v^{\gamma}(S)} + \sqrt{v^{\gamma}(T)} \leq \sqrt{v^{\gamma}(S \cup T)} + \sqrt{v^{\gamma}(S \cap T)} \tag{22}$$

Moreover, since $v^{\gamma}(S \cap T) \leq v^{\gamma}(S)$ and $v^{\gamma}(S \cap T) \leq v^{\gamma}(T)$, it follows from (22) that:

$$\left|\sqrt{v^{\gamma}(S)} - \sqrt{v^{\gamma}(T)}\right| \leq \left|\sqrt{v^{\gamma}(S \cup T)} - \sqrt{v^{\gamma}(S \cap T)}\right| \tag{23}$$

Squaring both (22) and (23) and combining gives us $v^{\gamma}(S) + v^{\gamma}(T) \leq v^{\gamma}(S \cup T) + v^{\gamma}(S \cap T)$, so $(N, v^{\gamma})$ is a convex game. By Shapley's theorem, the γ-core is non-empty. ∎